# Foundation Models for Astrobiology: Paper I – Workshop and Overview

Authors: Ryan Felton (NASA ARC), Caleb Scharf (NASA ARC), Stuart Bartlett (California Institute of Technology/SETI Institute), Nathalie A. Cabrol (SETI Institute), Victoria Da Poian (Tyto Athene LLC/NASA GSFC), Diana Gentry (NASA ARC), Jian Gong (University of Wyoming), Adrienne Hoarfrost (University of Georgia), Manil Maskey (NASA MSFC), Floyd Nichols (Virginia Tech), Conor A. Nixon (NASA GSFC), Tejas Panambur (University of Massachusetts, Amherst), Joseph Pasterski (NASA GSFC), Anton S. Petrov (Georgia Institute of Technology), Anirudh Prabhu (Carnegie Science), Brenda Thomson (Rensselaer Polytechnic Institute), Hamed Valizadegan (NASA ARC / KBR), Kimberley Warren-Rhodes (SETI Institute/NASA ARC), David Wettergreen (Carnegie Mellon University), Michael L. Wong (Carnegie Science), Anastasia Yanchilina (SETI Institute).

**Running title:** Foundation Models for Astrobiology

**Corresponding author:** Ryan Felton (ryan.c.felton@nasa.gov)

**Abstract:**
Advances in machine learning over the past decade have resulted in a proliferation of algorithmic applications for encoding, characterizing, and acting on complex data that may contain many high dimensional features. Recently, the emergence of deep-learning models trained across very large datasets has created a new paradigm for machine learning in the form of Foundation Models. Foundation Models are programs trained on very large and broad datasets with an extensive number of parameters. Once built, these extremely powerful, and flexible, models can be utilized in less resource-intensive ways to build many different, downstream applications that can integrate previously disparate, multimodal data. The development of these applications can be done rapidly and with a much lower demand for machine learning expertise. Additionally the necessary infrastructure and models themselves are already being established within agencies such as NASA and ESA. At NASA this work is across several divisions of the Science Mission Directorate. Examples include the NASA Goddard and INDUS Large Language Models and the Prithvi Geospatial Foundation Model. And ESA initiatives to bring Foundation Models to Earth observations has led to the development of TerraMind. A workshop was held by the NASA Ames Research Center and the SETI Institute, in February 2025, to investigate the potential of Foundation Models for astrobiological research and to determine what steps would be needed to build and utilize such a model or models. This paper shares the findings and recommendations of that workshop, and describes clear near-term, and future opportunities in the development of a Foundation Model (or Models) for astrobiology applications. These applications would include a biosignature, or life characterization, task, a mission development and operations task, and a natural language task for integrating and supporting astrobiology research needs.

## 1. Introduction

The search for evidence of life elsewhere in the universe remains astrobiology's central goal (National Academies of Sciences, Engineering, and Medicine, 2023). Living systems are thought to emerge as a dynamic interplay of non-equilibrium processes operating across multiple scales of complexity, dimensionality, and time. Consequently, to recognize life’s manifestations in unexpected contexts and forms (as might exist here on Earth or elsewhere), it makes sense to consider life as a complex multimodal, multidimensional universal process rather than a single state as it is commonly thought of on Earth.

The multidimensionality of life – meaning the many attributes or variables that define a living system’s properties – becomes apparent when we consider its simultaneous operation across scales (Figure 1). At the atomic and molecular level processes involve (or ‘exploit’) fundamental attributes of matter, including, for instance, quantum tunneling effects that facilitate efficient enzymatic functions (Bothma et al, 2010). At cellular scales, emergent networks of chemical reactions establish autocatalytic systems that maintain physical boundaries while remaining thermodynamically open. At organismal and ecological levels, complex feedback loops create resilient, adaptive structures that persist through environmental fluctuations. Finally, at the planetary level, life is revealed by mutualistic co-evolutions of geo- and bio-chemistries. Each scale represents a different expression of the phenomenon we call “life as we know it on Earth”.

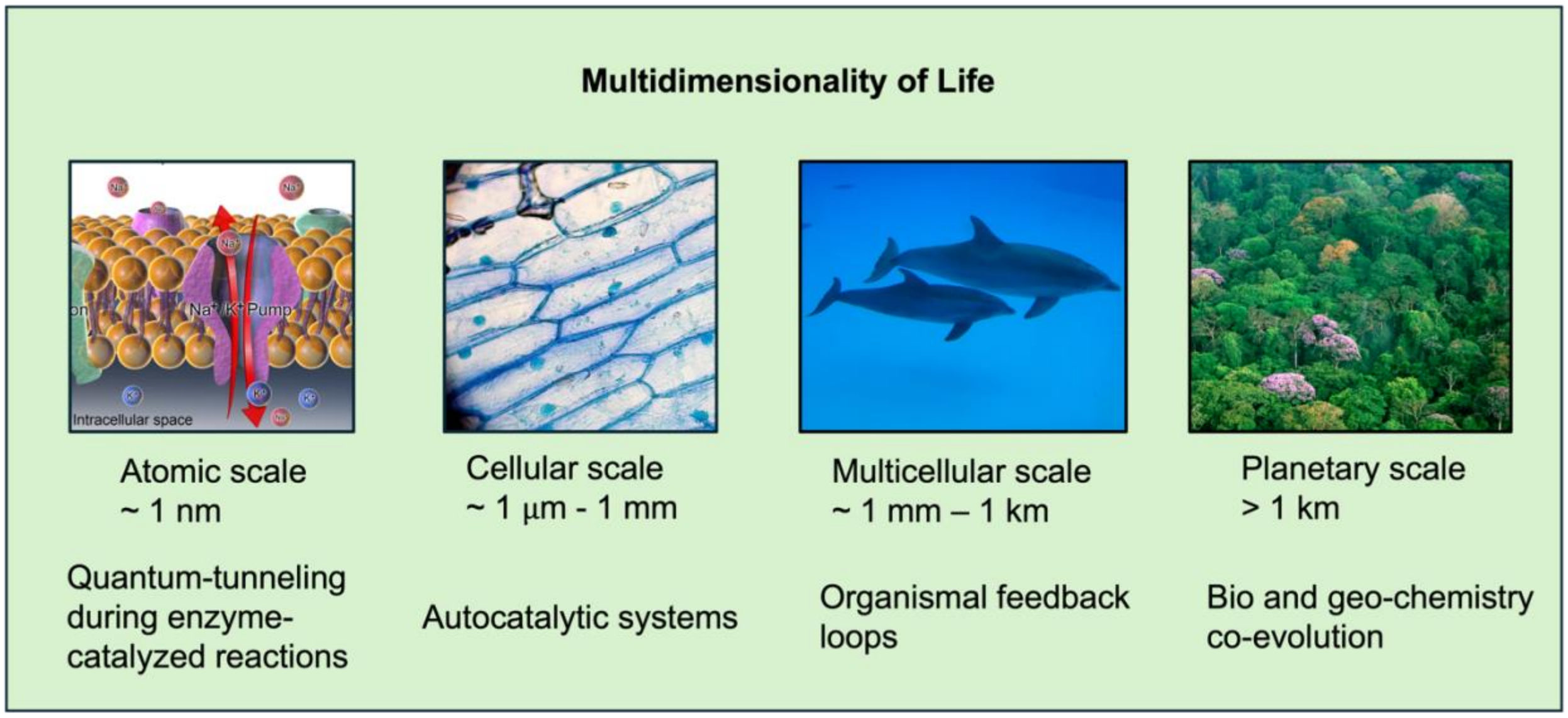

**Figure 1**: *The processes of life are evident at all size scales, from atomic to planetary. Image credits: Atomic scale Blausen Medical CC BY 3.0, Cellular scale Umberto Salvagnin CC BY 2.0, Multicellular scale Wikipedia public domain, Planetary scale Christian Ziegler CC BY 2.5.*

As a result, astrobiology research draws on a corpus of diverse multimodal, multiscale data that can characterize the subtle and high-dimensional features of living systems commonly termed biosignatures (e.g.,Des Marais et al, 2008). The effort to identify and interpret biosignatures is therefore one field of study where emerging artificial intelligence (AI), and in particular machine learning (ML) approaches, may be useful (Scharf et al, 2024). While AI is the overall field of using computer programs that mimic human intelligence, ML is a subset focused on computational techniques to discover patterns and make predictions. AI/ML can be leveraged to aid in identifying and interpreting biosignatures from complex and high-dimensional data. More specifically, AI/ML can identify a low-dimensional representation (latent space) of such data which may encode fundamental properties of life. Existing AI applications in astrobiology have ranged from examples such as the study of molecular complexity (Cleaves et al, 2023) and environmental biosignatures (Warren-Rhodes et al, 2023), to exoplanetary detection and analysis of planetary atmospheres (Valizadegan et al, 2022; Gharib-Nezhad et al, 2025; Cobb et al, 2019]).

As the amount of astrobiologically relevant data increases through field, laboratory, experimental, and simulation studies, AI approaches will become increasingly important for integrating different datasets and extracting the most deterministic evidence for life (Theiling et al, 2022). Likewise, a wealth of disparate, multi-scale, multi-modal astrobiological datastreams already exist that are immediately available for interrogation, integration and analysis with currently available and future AI/ML tools and models (Table 2).

Key strengths of AI/ML approaches include the development of frameworks to automate complex analysis of very large datasets and the ability to provide agnostic and potentially unforeseen insights into the underlying patterns in astrobiologically-relevant data. These frameworks may include machine classification, pattern or anomaly detection, and the recognition of non-linear, multiparameter and/or multicollinear properties in data, as well as generative modeling or simulation by learning. Recently, a particularly important ML concept, called Foundation Models, has emerged (FMs, see Section 3 below), which are large-scale ML systems typically trained on enormous datasets to encode informative data features and relationships. This enables FMs to serve as a foundation for various downstream applications, sometimes with relatively little additional data needed to achieve high performance (Bommasani et al, 2021). FMs go beyond many traditional ML applications in being able to handle larger amounts of diverse (i.e., multimodal) data while teasing out results from more specific and challenging interrelated subsets of data. An example of a type of FM becoming ubiquitous amongst society are Large Language models (LLMS). These FMs are trained on massive amounts of publicly available datasets and able to generate text, audio, and visual output.

To assess the potential of FMs for Astrobiology, the NASA Topical Workshops, Symposiums, and Conferences (TWSC) program funded a workshop held in February 2025 (see Appendix I for details of workshop participation, demographics, format, and materials). The overarching goals of the workshop were to: A) delineate the potential options and capabilities for astrobiology FMs (i.e., which topic areas and what data would be relevant); B) examine and outline what would be needed to produce FMs and what use cases might look like; C) create a short-list of potential FMs (or a single FM framework) with a focus on need, timeliness, current technological and data capabilities, expertise within the community, and future innovation; and D) bring the recommendations and findings of the workshop to the attention of the astrobiology community.

The overarching conclusion of the workshop (elaborated on in the rest of this paper) is a recommendation for **the development of a multimodal Astrobiology Foundation Model that focuses on the detection of life, mission-focused decision making, and is interfaced by an interdisciplinary text-based large language model (Figure 2)**.

The relevance of the workshop and findings to NASA and community strategic goals is highlighted below. Sections 2 and 3 expand on the challenges specific to current astrobiology research, and how FMs can address these challenges, and provide more technical details on FMs. Section 4 discusses the workshop, its general structure, and the path to these findings. Sections 5.1, 5.2, and 5.3 go into greater detail on the three principal workshop findings; sketching out the science, justifications, and roadmaps for each of them. Section 6 concludes with summaries of the key needs and options for FMs and AI applications for astrobiology research. These include comments on data coherence and the astrobiology data ecosystem, and a broader roadmap is presented for bringing an FM for Astrobiology into use in the near term and further term.

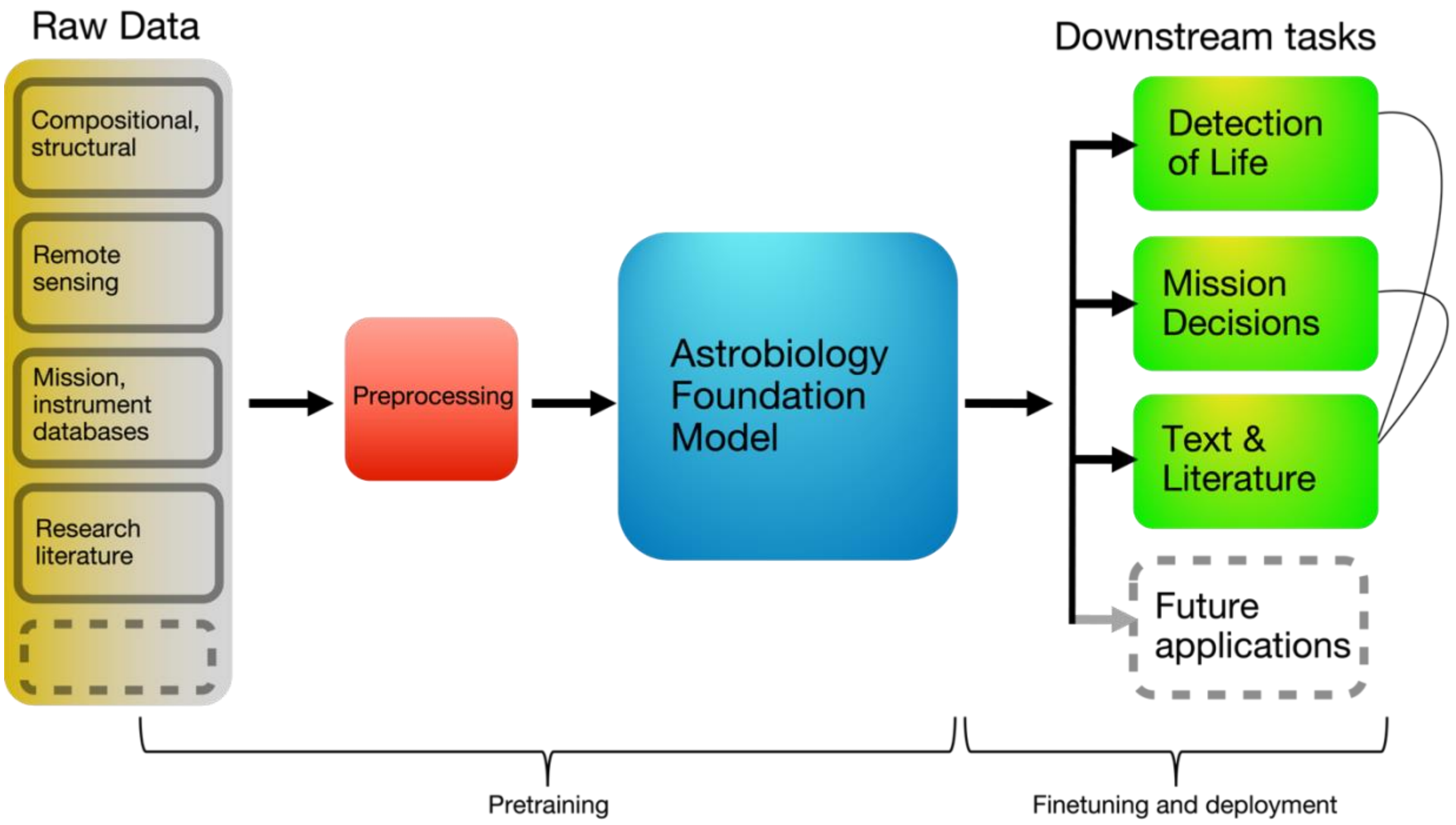

**Figure 2:** *Illustration of a potential multimodal Foundation Model for astrobiology, capable of a variety of downstream use-cases. For example: life detection, mission decision-making, and a natural language application for research and for engaging with the other applications.* The development of such an FM would be supported by an initial

period of self-supervised/semi-supervised pretraining and then fine-tuning and public deployment of the downstream tasks/applications. After deployment, optimization would occur based on the public's use and input (via a performance feedback improvement system), and integration of new data sets as they become available.

**Relevance to NASA's strategic goals**: The workshop and its findings are directly relevant to the NASA Science Mission Directorate (SMD) Planetary Sciences programs, and Astrobiology, as well as SMD's Science-enabling Technology, and are responsive to multiple NASA-related reports, and recommendations. In Table 1 the findings are highlighted together with their relevance to recommendations #5 and #11 from SMD's Strategy for Data Management and Computing for Groundbreaking Science 2019-2024 report and the Origins, Worlds, and Life (OWL): A Decadal Strategy for Planetary Science and Astrobiology 2023-2032 report (National Academies of Sciences, Engineering, and Medicine, 2023).

| Report | Goal/ Objective/ Recommendation | Details | Workshop Findings Relevance |
|---|---|---|---|
| SMD Strategy for Data Management and Computing for Groundbreaking Science 2019-2024 | #5 | *"[...]strongly encourage collaboration and cooperation of data professionals in academia, industry, and elsewhere to enable cross-cutting scientific discovery."* | Our participants are a mix of astrobiologists and data scientists. We deem this collaboration a necessity to fully grasp the challenges and solutions to bringing FM's to fruition. |
| SMD Strategy for Data Management and Computing for Groundbreaking Science 2019-2024 | #11 | *"[...] to incentivize and educate the community on how to use AI/ML to approach science in new ways[...]"* | FM's provide a novel AI/ML approach to astrobiology that can be applied to multiple analysis techniques/tools (i.e. mass spec, GCM's) |
| Origins, Worlds, and Life: A Decadal Strategy for Planetary Science and Astrobiology 2023-2032 | General Technology Area - *Autonomy* | *"[...]execute planned operations on remote yet complex planetary science and astrobiology missions. Machine learning/artificial intelligence can support the implementation of autonomy in such environments."* | A potential FM application is in situ decision making and execution during planetary missions. |

**Table 1**: *Highlighted relevance between NASA strategic report goals and recommendations, and the FM workshop findings.*

## 2. Astrobiological data challenges and machine learning challenges

Astrobiological data streams (e.g., Figure 2) are often expensive to produce and require domain knowledge expertise to construct and utilize. As a result, numerous astrobiology datasets observe only a small number of data points, and can be very sparse, as well as geospatially and temporally narrow. Furthermore, different data modalities relevant to mission instrumentation (Table 2) are seldom integrated and even more rarely combined with laboratory studies and measurements from terrestrial analogues.

| Discipline | Data Types/Streams |
|---|---|
| Geomorphology | Landscape imagery, high resolution topography |
| Remote Sensing | Mineralogy, elemental abundance composition |
| Aqueous Geochemistry | Physicochemical composition |
| Geomicrobiology | Genomics |
| Organic Geochemistry | Lipid biomarkers |

**Table 2**: Some *examples of typical astrobiology data modalities. Geomorphology (Palucis et al, 2014; Rogers et al, 2023), Remote Sensing (Warren-Rhodes et al, 2023; Nichols et al, 2024; Harris et al, 2022), Aqueous Geochemistry (Tosca et al, 2008), Geomicrobiology (Li et al, 2023; Pontefract et al, 2017), Organic Geochemistry (Wilhelm et al, 2017; Nichols et al, 2023; Georgiou and Deamer, 2014).*

The data indicated in Table 2 have different formats and format standards, spatial scales and resolution, time scales, variety/types, and are rarely combined when traditional analysis methods are utilized. While there are a few community-recognized data repositories and proposed metadata standards (such as the Astrobiology Habitable Environments Database - AHED, https://ahed.nasa.gov/, and Wolfe et al, 2024). There are also very few studies which have used measurements from multiple diverse datasets with applications to astrobiology (Warren-Rhodes et al, 2023). Because the confidence in the detection, identification, and characterization of any potential extraterrestrial life should – in principle – increase with multiple combined measurements (Neveu et al, 2018), and because any potential extraterrestrial life detection would have to be subject to the highest level of scrutiny, integrating the informational features contained in multiple measurements is essential to eventually confirm the detection of extraterrestrial life.

In that context, ML in general, and FMs in particular (see Section 3 below) can help optimize and streamline the analytical pipeline needed to process and integrate the complex multimodal datasets required for astrobiology. The availability of sufficiently large unimodal data streams relevant to astrobiology is limited given the data sizes typically required for useful deep ML model performance, and this is especially true for multimodal data streams which are even more rarely gathered into a single data corpus. The task of detecting or predicting habitability and potential biological phenomena beyond Earth therefore still poses significant challenges to AI/ML that will require careful data processing, model development, design, and tuning, and training strategies to best capture astrobiology-relevant features.

Furthermore, extraterrestrial measured conditions will inevitably be novel, with a distinct distribution relative to the range of observable conditions on Earth, requiring AI/ML models to generalize to these out-of-boundary settings. Crucially, any predictions made by a model trained on Earth-based samples must extrapolate and be adaptable in case of highly novel and unique environments that might be encountered on other worlds, representing the "life as we don't know it". Building generalizable FMs that extrapolate to astrobiology-relevant conditions likely requires evaluation criteria and datasets that are distinct from those typically used in ML and may come at the expense of predictive performance when applied in Earth-based environments (i.e., the capacity to generalize may reduce efficacy for Earth-specific conditions). Model developers should be mindful of this need for adaptation and the implications for the design of validation/test evaluation sets and the use of modeling approaches to minimize overfitting (i.e., reduced efficacy at generalizing to unseen data) in order to provide realistic evaluations of the degree of generalization expected during deployment. Model architectures and training approaches that improve performance in few-shot (where very small, labeled training data are utilized), and zero-shot (where a model generalizes to new, unseen classes) settings are paramount, given the small datasets in the field of astrobiology. In addition, these challenges will provide the opportunity to assess the limitations of an FM and assist in refinement of the model.

## 3. Foundation Models

To recap, an FM is defined as a large-scale ML system typically trained on enormous datasets to encode fundamental, general information and relationships, enabling it to serve as a foundation for various downstream applications with limited additional training and fine-tuning (Bommasani et al, 2021). The encoded information can be in the form of complex correlations and patterns that, when learned, enable fast and sophisticated queries and analysis of the original data and – critically – rapid refinement of the model for specific, specialized tasks without the cost of the initial training.

The operation of FMs can also be thought of, and used, as a type of data compression that not only preserves the meaningful information but structures it in a way that is optimal for its further use. Large language models (LLMs) are the most well-known example of FMs, and are used downstream to create chatbots, software coding assistants, and 'experts', and so on, that focus on language as a primary data source for input and output. FMs can be fine-tuned on smaller, bespoke datasets (e.g., the results of a specific investigation or use of a new measurement instrument) to provide more accurate and relevant analyses by leveraging the prior knowledge encoded in the model's internal data representation.

While building such large-scale FMs requires a great deal of ML expertise, effort, and resources, adapting them for specific downstream tasks is often fast and requires ML expertise to a lower extent, making them excellent tools for rapid development in fields that call for a wide range of applications. In the last few years, FMs have emerged as a new paradigm that can dramatically accelerate the application of ML to specialized tasks and a wider range and types of data (Szwarcman et al, 2024; Shinde et al, 2024). For astrobiology, FMs may offer unique and critical opportunities for advancing efforts in life detection and characterization. For example, FMs may be useful in extracting any information (in features within data) that could be interpreted as an outlier to an abiotic baseline which itself might only be properly quantified following analysis of large, federated datasets or combinations of disparate variables not intuited within traditional disciplines (Section 5.1).

This versatility makes FMs increasingly important in a number of scientific areas, including molecular biology (protein folding/prediction; Nussinov et al, 2022), metagenomics (Mathieu et al, 2022; Tonkovic et al, 2020), microbial ecology (Warren-Rhodes et al, 2023; Roussel and Bohm, 2023), and climate science and modeling (Mukkavilli et al, 2023) for which the same datasets are used for many different tasks. Recently, NASA and IBM have built the Harmonized Landsat and Sentinel-2 (HLS) Geospatial FM that uses the extensive HLS imaging data of the Earth to perform geospatial analysis (Jakubik et al, 2023). This FM encodes informative features in an internal (latent) representation of the Earth's surface that is broad in scope, but captures all of the correlative, patterned relationships of land and ocean composition, optical properties, topography and so on. Subsequently, once trained, the FM (which is described entirely by its many internal parameters) can be quickly and easily replicated and fine-tuned for inference tasks such as evaluating fire zone risks or flood risks from higher resolution or newer data. These capabilities, and those of many FMs, are made possible by advances in deep learning—particularly self-attention mechanisms like those used in Transformer architectures—which allow models to identify and encode semantic relationships in data. This enables FMs to build internal representations that are not only comprehensive but also highly transferable across diverse tasks.

FMs are beginning to be studied in areas even more directly adjacent to astrobiology. For instance, an astronomical FM has been proposed and built for stellar astronomy (Leung and Bovy, 2024) that can be applied to data such as that of the Gaia mission to generate spectra from stellar parameters and perform a variety of discriminatory and inference tasks that include predicting interstellar extinction curves and filling gaps in spectra. ClimaX, has been developed as a foundation model specific for weather and climate prediction which allows model and forecast, weather and climate projections (Nguyen et al, 2023). An example of a mission with data sets that span both the unimodal and multimodal environment and the divisional space of astrophysics and astrobiology is the brand new Vera Rubin Observatory (Vera C. Rubin Observatory, 2025). During its 10 year Legacy Survey of Space and Time, Rubin will be producing exceptionally large amounts of image data at various scales and lead to follow-up spectroscopic studies. This type of data ecosystem would benefit from the use of an FM to help coordinate and prioritize the various multi-scale studies.

Critically, FMs do *not necessarily* require huge datasets (although they may be more limited in scope as a consequence), and they can be explicitly 'multimodal' by utilizing disparate datasets. Furthermore, more than one form of FM can be combined for new tasks across different types of data. For example, a protein–large-language model (that captures information about protein functional syntaxes) can be combined with a model like AlphaFold (that captures the physics of protein folding) and an LLM to provide a natural language interface to analyzing protein structure and function (Xiao et al, 2024).

However, as with all analytical techniques in science, and particularly AI/ML methods, FMs present numerous challenges in reproducibility and generalizability of results and behaviors, traceability of cause-and-effect in data, and bias due to either explicit decisions in data-labeling or hidden choices in unlabeled data at the very outset of data acquisition or even instrument or experiment design. For astrobiology, where life detection is a central challenge, all such issues may be especially important to characterizing any FM or its downstream, fine-tuned, application. This is particularly critical since FMs could also be leveraged not just for life detection but for habitability assessments following characterization of local micro-environments during mission planning and exploration (e.g., Section 5.2). To instill confidence in an FM and its associated multimodal modules, we propose a comprehensive validation and benchmarking framework closely aligned with best practices from foundational model research. This framework is modular in design; each modality first undergoes targeted benchmarking (e.g., hypothesis reconstruction from mission outputs, spectral interpretation accuracy, domain specific omission robustness). Then, so that each modality can work with another modality (cross-modal processes) the output from the linking (cross-modal integration) is separately evaluated for consistency and collaborative synergy. Ground-truth tasks, defined a priori by domain experts, serve as anchors. Human baselines and expert review panels evaluate interpretability, hypothesis plausibleness, citation accuracy, and epistemic reliability, providing transparent comparisons to model outputs. To avoid overfitting and benchmark leakage, protocols rely on reproducible data splits, clearly documented hyperparameter settings, and statistically representative sampling. Robustness is tested through adversarial and out-of-distribution queries that probe interpretive limits. This evaluation is embedded in a performance assessment feedback loop, enabling iterative refinement of the system to reinforce the human-centered quality-control cycle.

## 4. The Workshop:

The Foundation Models for Astrobiology Workshop was structured as a two-part event composed of a one-day virtual primer followed by a three-day in-person component. Due to the workshop participants' backgrounds across a diverse set of disciplines spanning data science and astrobiology, a series of online talks was given beforehand to ground participants in a common set of interdisciplinary knowledge (see Appendix). This allowed participants to establish a baseline of understanding across the disparate fields and to facilitate discussion. This virtual primer covered: fundamentals of machine learning, foundation models, best practices of AI/ML use, fundamentals of biosignatures, exoplanet characterization and habitability, in-situ autonomy technology, and astrobiological sample composition.

The in-person portion of the workshop took place at the SETI Institute in Mountain View California from February 24–26, 2025. To help sculpt initial discussions, representatives from all of the NASA Astrobiology Research Coordination Networks (RCNs) presented introductions on the RCNs and discussed current research priorities and instances where these pertained to AI/ML. The RCNs are: the Nexus for Exoplanet System Science (NExSS), the Network for Life Detection (NfoLD), the Network for Ocean Worlds (NOW), LIFE: Early Cells to Multicellularity, and the Prebiotic Chemistry and Early Earth Environments Consortium (PCE3).

The workshop then created a list of wants and needs for astrobiology that AI/ML (and FMs in particular) might be able to support. A set of themes and critical questions quickly emerged around three initial categories: patterns and interpretations, missions, and hypothesis driven science. Breakout groups revolving around these categories were formed and the categories were discussed in detail.

The patterns and interpretations group homed in on questions involving patterns and fundamental relationships when attempting to identify biotic versus abiotic samples. This eventually led to discussion of a Life/Non-life boundary and its contrast with the idea of a continuum between living and non-living systems, with the group identifying the need to develop an ML application for biosignature detection and the potential for an FM to serve as the backbone application for this work (Section 5.1).

The mission group's themes and questions were focused on mission design and mission operations and how AI/ML could augment these tasks. For example, envisioning the use of AI/ML to supply answers to a query such as: *"Given the current measurements of a solar system body, what location on it has the highest confidence interval of having detectable biosignatures?"* This led to the second finding of the workshop: the need for AI/ML involved in all stages of mission design and operations and an incentive to build an Astrobiology Mission Model (AMM) to accomplish these tasks (Section 5.2 below).

The third and final breakout group, "hypothesis," was centered around the idea of utilizing AI/ML to assist scientists during their research. After discussion of a variety of use case situations, including hypothesis generation, this group came to the conclusion that a text-based model trained on the corpus of research literature, mission documentation, and adjacent materials, has significant potential. A hypothetical "Astrobiology-Chat" or AB-Chat LLM could support both research needs and communication needs and be integrated with other downstream instances of a FM (see Section 5.3 below).

**The overarching conclusion of the workshop (Section 6) was that a multimodal FM for astrobiology has great potential**. In addition to advancing specific science goals (e.g., life detection) such a model would call for the advancement of the astrobiology data ecosystem and would align with both robotic and human exploration goals. Equally, the specific architecture of an astrobiology FM requires further study. While a multimodal model is indicated, other options include a set of unimodal FMs built around different model structures (e.g., GPT versus BERT architectures for natural language processing).

In the following sections these findings are described in detail, based around the three focus areas/use cases identified in the workshop.

## 5. The Findings:

### 5.1 Detection of Life

Defining the essential properties of life, and criteria for evidence of life, are two views of the same problem. While many past efforts have been made at universal definitions of life (as we know it) from first principles, none have convincingly resolved all outstanding problems (Neveu et al. 2018; Cleland 2019). This difficulty stems from at least two sources. One is the present lack of an example of non-terrestrial life; distinct biochemistries may be produced by the coevolution between alien life and non-terrestrial environments that may demand detection strategies that move beyond Earth-based assumptions. The other is that the distinction between biotic and abiotic processes may not be clearly delineated (Figure 3) but rather present as a context-dependent continuum (Jheeta et al, 2021; Ratliff et al, 2023).

Within each currently explored biosignature modality (from, for example, isotopic abundances to molecular species or genomics), available training data typically encompasses only a narrow band of biological and abiotic parameters—with the former constrained to well-characterized extant terrestrial biology. For example, these datasets lack crucial information on various emerging states of matter, such as ancestral geochemical states, prebiotic systems, or the earliest signs of co-evolution. Moreover, they mostly provide no representation of potential alternative biochemistries (see however, Chandru et al, 2024). This incomplete and Earth-centric coverage could severely limit our ability to generalize biosignature detection methods, especially for life as we do not yet know it.

It is proposed that AI/ML, with the capacity to handle multi-dimensional data (see Section 1 above) might help overcome these limitations (at least within the known landscape of complex chemistries) and enhance the ability to envision, predict and/or detect extraterrestrial life across its potentially diverse manifestations. A major opportunity would therefore come from creating a FM that integrates datasets across different modalities (e.g. informational, molecular, structural, metabolic, coevolutionary) to establish a reliable framework of integrated biosignatures. This approach would enhance confidence in detecting extraterrestrial life, even when individual, unimodal, signatures remain biased and incomplete.

The ultimate goal would be to establish the appropriate mechanisms that integrate various modalities and training datasets into a unified representation space where correlations across modalities become apparent. This integration layer must be sophisticated enough to weight different modalities according to their reliability and relevance in specific contexts, addressing the problem of incomplete or biased individual signatures.

Emerging AI/ML tools (Cleaves et al, 2023) offer a potential new approach not only towards identifying complex combinations of data signatures that could be diagnostic of life, but also towards revealing *which* combinations are diagnostic (and even improve our understanding of what life on Earth is) – in essence, an empirical definition of life. **It is therefore suggested that, as part of an initial development stage and application of an astrobiology FM, a multimodal abiotic and biotic database be built with the specific purpose of biosignature detection as a downstream task**. Building that database and a FM for biosignature detection will require an iterative approach, starting with proof-of-concept validation using existing data, refining its predictive power by addressing key knowledge gaps, and ultimately integrating diverse datasets from missions and terrestrial biosphere studies to create a scalable, adaptive framework for life detection.

Work like this would address a number of outstanding questions. For example, what data is the most valuable in training an FM or its downstream use cases for life detection? One approach might be to train multiple FMs (if sufficient data existed), using different combinations of data inputs, and fine tuning those FMs to predict biosignatures - in order to identify which data inputs are actually necessary to get best performance. An alternative approach to achieving the same goal would be to train one FM with all the data inputs, and perform feature importance analysis, with pruning or explainable AI approaches. Or, many unimodal model FMs might be trained and combined into multiple multimodal FMs with adapters or contrastive learning etc. Overall this work would allow the community to tackle important life and non-life data-driven questions (Figure 3). *What are the optimal features/dimensions for distinguishing life and non-life? Can we identify potentially anomalous biosignatures with unsupervised approaches? Is there an empirical, multidimensional boundary/gradient between life and non-life, and if so, what is it?* Dimensionality reduction and clustering are just some of the unsupervised machine learning methods that assist in finding hidden structures and patterns. In all such cases a broader issue is how to evaluate such (fine tuned) FM models to decide which ones are more predictive than another.

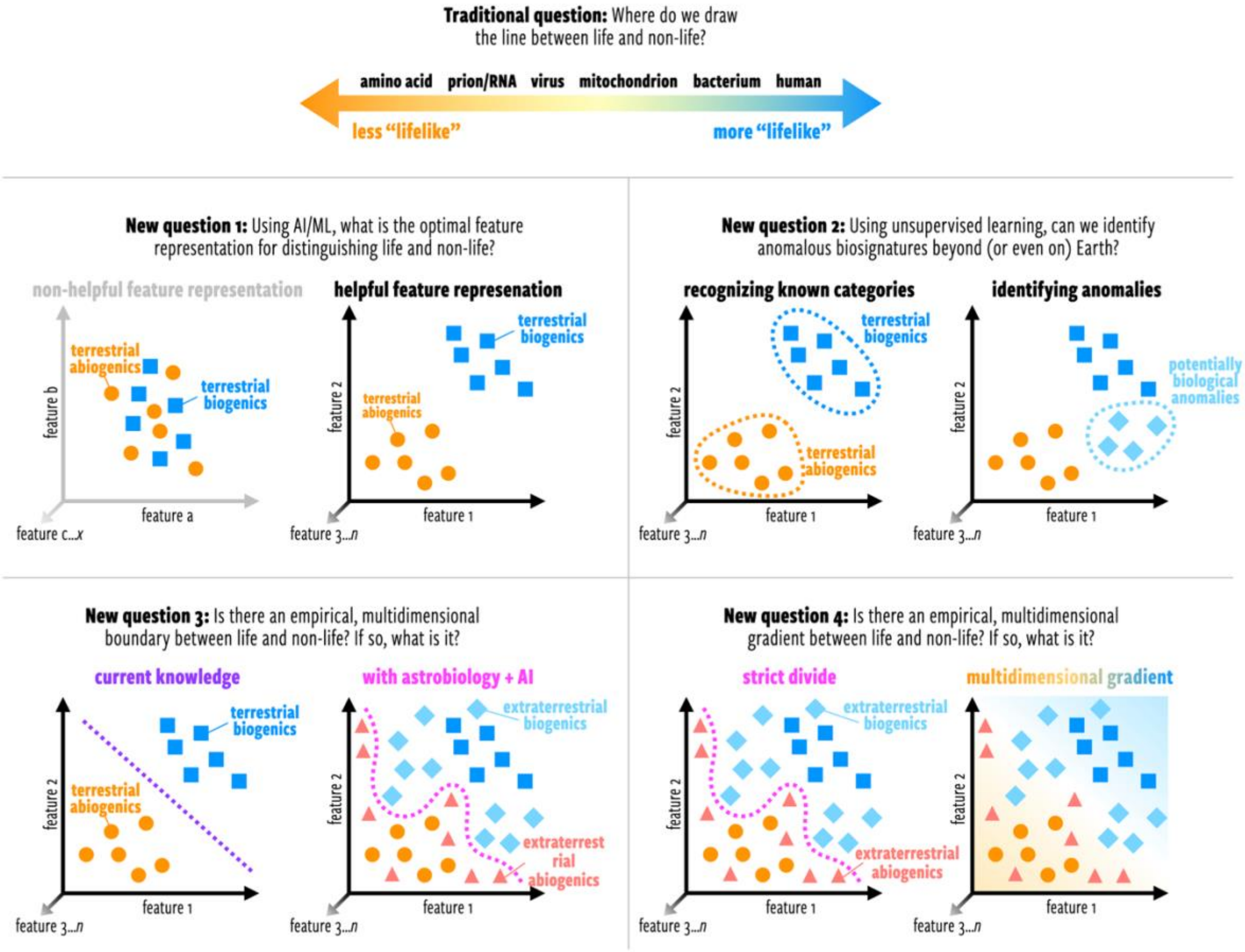

**Figure 3.** *AI/ML methods, and FMs in particular, may allow us to advance beyond the ago-old astrobiological question, “Where do we draw the line between life and non-life” to more data-driven questions.*

**Requirements for developing a “detection of life” FM use case:**

**(1). Developing a proof-of-concept by revisiting mission data, experimental work, and terrestrial analogs.** The development of a robust FM leading to a downstream task for biosignature detection begins by leveraging existing datasets to construct an initial proof-of-concept, which may include comparing similar datasets between terrestrial analogs/Mars and the moon/asteroids as “abiotic” endmembers. Existing datasets include AHED (organics and mineralogy), the Planetary Data System (PDS, satellite imagery and reflectance data, https://pds.nasa.gov), Sedimentary Geochemistry and Paleoenvironments Project (SGP, sedimentary geochemistry, https://sgp.stanford.edu), pyrolysis-gas chromatography-mass spectrometry data repositories (Cleaves et al, 2023), and Lunar Reconnaissance Orbiter Camera (LROC) imagery (Lesnikowski et al, 2024). Several data types are identified where it is believed that there is sufficient breadth in the literature to use in the development of the proof-of-concept. These include visible imagery, VNIR reflectance, elemental and isotopic abundance, GC-MS, Raman, XRF/XRD, and topography. This subset represents low-hanging fruit, with the potential to add more data types depending on the progress of open science initiatives and data standardization. The next steps will then be extending the model by:

**(2). Integrating datasets and closing knowledge gaps.** An integrated approach will allow for flexible, biosignature detection, allowing the creation of an iterative, scalable framework that adapts to new discoveries. One objective would be to synchronize data collection across planetary missions and terrestrial datasets, ensuring that all data are based on the same measurements (e.g. compositional, structural, meteorological) and are standardized. Another objective would be to address key data gaps, particularly those that are most urgent in the near term and directly relevant to upcoming astrobiology missions (e.g., progress in distinguishing biotic from abiotic signatures, and mimics). Filling these gaps will (i) strengthen a model's predictive power and enhance its ability to recognize coevolutions beyond Earth (i.e., biotic, abiotic, mimic signatures generated by a world's specific physical and biological constraints); (ii) leverage the multimodal nature of the FM.

To ensure the validity of such an FM (or FM use case), multiple layers of measurement/data must be combined, which may lead to new insights, correlations and patterns. These combinations might revolve around:

- Chemical and morphological correlations
- Biological and ecological relationships
- Spatiotemporal patterns
- Contextual analysis (required to evaluate potential biosignatures within their planetary environment)
- Cross-validation through multiple independent detection methods

By unifying these elements into a multimodal detection framework, this model would represent the most comprehensive and adaptable strategy for identifying potential biosignatures and ruling out abiotic alternatives, offering the most effective path forward in the search for life beyond Earth, and adding depth to our understanding of life on Earth.

**(3). Generalized and targeted exploration strategies.** Expanding application tasks alongside this effort would ensure that a model evolves with real-world mission constraints in mind. The goal would be to synchronize data collection across planetary missions and terrestrial analog works, ensuring that all mission and analog data are compatible and complementary. This integrated approach would enable flexible, adaptable biosignature detection, allowing the creation of a scalable framework that adjusts dynamically to new discoveries. A truly comprehensive FM would require all available data (chemistry, geology, and biology) to maximize the potential for detecting life beyond Earth. An ultimate advancement would be for the FM (or FM use case) to be capable of encoding or formulating the basic principles of life and help to refine the concepts of biosignatures, abiotic and mimic materials and processes, as well as to develop targeted exploration strategies (e.g., science questions, hypotheses, experiments, instruments, technology*)* for specific worlds.

### 5.2: Mission-focused Decision Making

Analyses performed on other planetary bodies provide direct assessments of extraterrestrial environments making flight missions extremely valuable for detecting non-Earth-based life forms and their associated processes. However, flight missions are resource- and time-constrained compared to terrestrial laboratory analyses due to size, weight, and power (SWaP) limitations, as well as communication delays and data transfer rate limitations. As missions are sent further into the solar system including the Juice mission to explore the icy moons of Jupiter (Fletcher et al, 2023), Europa Clipper (Pappalardo et al, 2024), and the Dragonfly mission to Titan (Barnes et al, 2021), the constraints of flight missions will become amplified. Communication, power draw, data compression, mass, and the ability to respond to data for real-time sample prioritization are just a few of the parameters that become increasingly difficult to balance with mission needs and provide impetus for greater autonomous and predictive analytical and decision making power. **We therefore suggest that the second downstream application of an astrobiology FM should be a specialized Astrobiology Mission Model (AMM) to maximize the scientific return of all planetary missions.** The AMM would be utilized during both the mission design and mission operations phases to limit the potential for inconclusive measurements and increase the likelihood of detecting extraterrestrial life via *in situ* analysis.

**AMM Applications**:

The AMM would be made available for mission design teams to fully optimize mission payloads during the design phase. A primary optimization goal would be to assess potential instrument suite limitations in a given planetary environment and determine additional measurements that would be needed to increase confidence in the interpretation of an astrobiologically relevant measurement in a specific site on a planetary body (Figure 4). Importantly, this application would be used to maximize the amount of information and resources available to the mission design team; while all information from the AMM would still be reviewed and cross-checked by subject matter experts. In other words, the AMM's function would be as a supplementary agent to assist human scientists, engineers, and project management members in mission design (Figure 4).

During mission operations, the AMM would use the exact specifications of the optimized instrument payload along with background knowledge of the specific location on the target planetary body to assist in decision making during time- and resource-constrained operations (Figure 4). The AMM could assist mission operations teams in assessing and synchronizing downlinked data and available terrestrial data to rapidly optimize subsequent measurements to meet mission science goals. Alternatively, in the longer-term, an onboard AMM could facilitate agentic AI during mission operations enabling autonomous decision making and real-time responses to *in situ* measurements (Figure 4; see Chien et al, 2024), even for missions in the outer solar system with relatively long data-transmission times (see Theiling et al. 2022). The degree of autonomy granted to the AMM would be curated by the mission team for the specific mission.

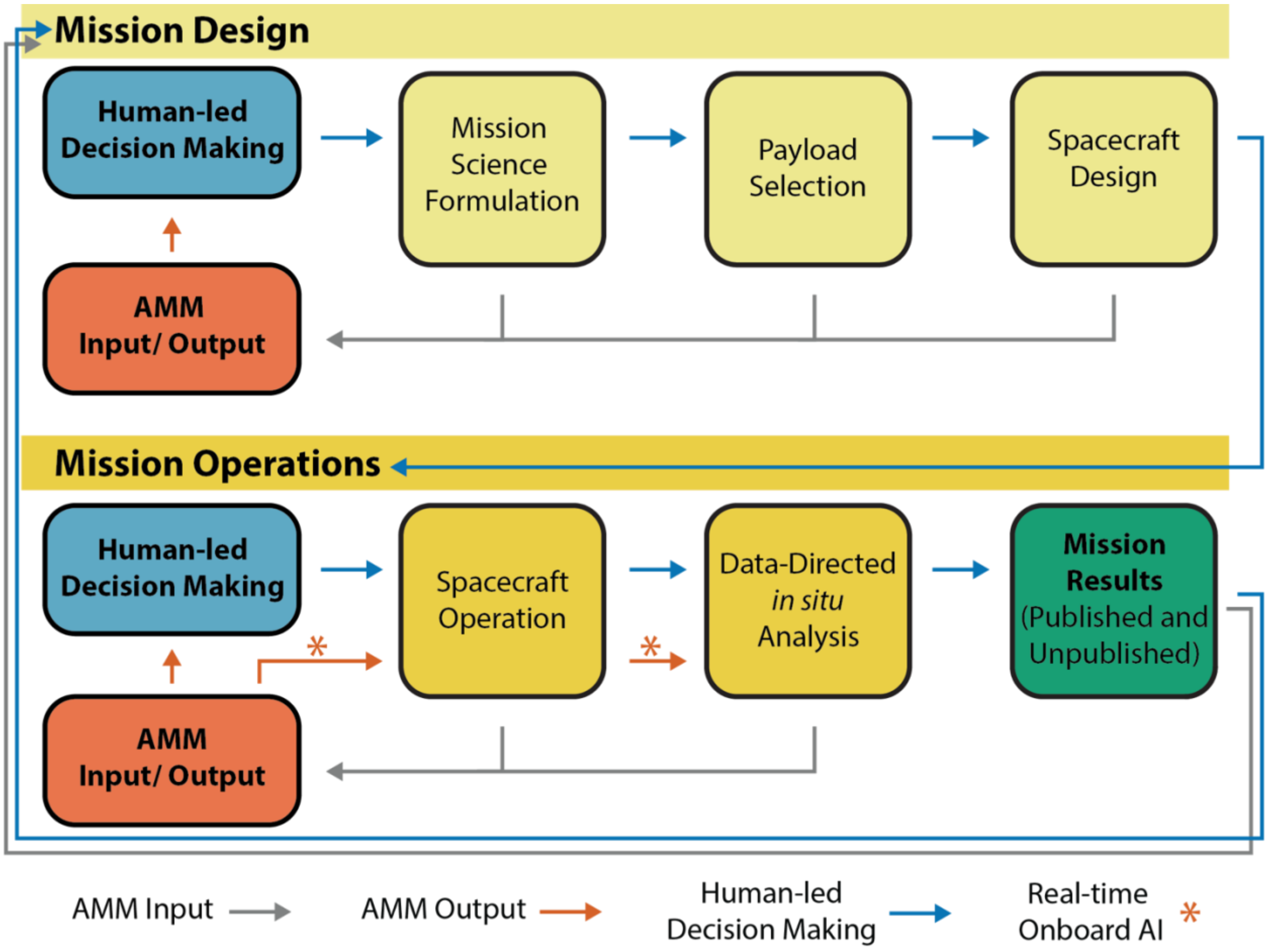

**Figure 4**. *Example of a potential human-AI partnership workflow during mission formulation, design, operations, and data analysis. Real-time decision making is made possible with onboard autonomous AI.*

The applications of the AMM would be extensive and not limited to this initial use scenario. If supplied with proper data, the AMM could assist and streamline common and often complex mission-related problem-solving, such as instrument malfunctions that may affect or cross multiple subsystems, or issues with calibration not recognized earlier in mission development. Such an AMM would also be valuable for evaluating datasets collected during previous missions, terrestrial analog campaigns, and laboratory studies to help identify patterns in the data and/or produce weighted probabilities that may indicate life, biotic or abiotic processes, or habitable environments, including those not detected during initial data interpretation. Furthermore, the AMM could be deployed in conjunction with other available technology to assist in sample triage and sample selection to locate targets with the highest likelihood for biosignature detection in conjunction with the mission team, potentially using augmented reality (AR). AR technology could be applied in either robotic or human-led exploration missions for tasks that require highly specific expertise that are enhanced with an increase in contextual information, including sample selection, site interpretation, or maintenance.

**AMM development approaches:**

For mission design, infrastructures already exist to begin developing an AMM (Verma et al, 2023; Vaquero et al, 2024; Bowkett et al, 2025), and to handle mission-relevant secure data, such as the LLM based ChatGSFC hosted at the NASA Goddard Space Flight Center (NASA IT Talk, April 2025; CAIO Team at GSFC, Personal Communication). To develop the AMM, a multimodal astrobiology FM would be fine-tuned with available database knowledge covering previous missions – including instrument suites, instrument capabilities, examples of the corresponding science traceability matrices (STMs) when available, measurements collected (e.g., mass spectrometry data, XRF data, etc., all of these experiments including metadata) as well as literature reporting potential problems with that data. Additionally, training on laboratory capabilities and analog field campaigns would be crucial for AMM tuning. This would include commercial and prototype instrument capabilities, previously developed proposals and mission concept studies, as well as available analog studies, including laws and regulations that may impact analog field campaigns. The collection of the data necessary to train the AMM could begin now, with additional data added during the development of the Astrobiology FM.

For mission operations, the AMM could be further augmented with specific details on the mission-specific instruments, including instrument capabilities, similar commercial analogs, relevant datasets produced on breadboard and engineering models for the mission, data transmission limitations, and mass and power constraints to optimize instrument use. Additionally, details about specific biosignatures potentially relevant to the selected planetary body and site that have not yet been incorporated into the AMM could be supplied (also synergistically with a detection of life use case). Analog studies could also be conducted during mission preparation to increase the performance and achieve the defined mission goals. Taken together, the incorporation of ML into mission design and operations is inevitable. An AMM could enhance mission operations during flight as well as science data interpretation, enabling a more successful mission from conception to operation that would maximize the chance of achieving mission objectives.

**Barriers to AMM adoption**

While much of the data needed for the development of the AMM exists, multiple barriers would need to be overcome. The most tangible barriers include the lack of available funding for the necessary work, paywall limitations from specific journals, the need to “tokenize” measurement data to make it transferable as inputs to the FM, the ML readiness of available data, and synchronizing the various datasets together. Additionally, all mission-sensitive data, including schematics, instrument specifications, internal testing procedural documentation, and data collected for the mission, could only be included on the AMM if it were hosted on a secure server.

Potential pathways to demonstrating the value of the AMM would include incorporating the lessons learned from missions either currently deploying or planning to deploy autonomy during mission operation (see Nesnas et al, 2021; Theiling et al, 2022; Verma et al, 2023). While there is subtlety in the distinction between autonomy and AI (autonomy focuses on task completion, AI often focuses on problem solving), the risk/reward-threshold to cede control of space missions to computers during operations has been crossed (Nesnas et al, 2021; Verma et al, 2023). As the need for such autonomy and AI increases for missions with diverse targets, a unified AMM-type architecture has the potential to increase the speed and reduce the cost and resource limitations (time, analytical power) of mission development. Notably, while hesitancy to adopt an AMM-type system can pose an understandable barrier to removing humans from the loop during mission operations, we have already seen the Mars rover program successfully increase the amount of autonomy in each new rover iteration, while still requiring human input and guidance in critical decisions (Verma et al, 2023). There is also the need to reinforce the view of an AMM as a tool to be used hand-in-hand with humans, not a device that removes them entirely from the process. Including humans-in-the-loop (HITL) with a deployed AMM could ensure that the AI agent is accurately interpreting the human-defined tasks and completing those tasks ethically and safely while still increasing the speed and efficiency of operation and science return of the mission relative to what could be accomplished using human operators alone. An AMM, even with HITL, would need to be extensively and rigorously tested prior to deployment. Pre-flight tests for the HITL-type AMM would need to push the extreme bounds of the AMM using accurate representations of the defined mission operations, just as any technology incorporated onto a flight mission.

It is suggested that an AMM should initially be used by mission development teams to increase the amount of information available, assist in recognizing patterns and connections not immediately available to the mission teams, as is typical for AI, and to ensure that baseline mission goals can be met with a proposed instrument suite. For missions developed with the AMM, it would be mandatory to adequately report the use of the AMM. Furthermore, for the foreseeable future all missions would still be reviewed and selected within the traditional, human-driven, mission review framework, regardless of AMM use. Additionally, in order to increase the robustness and trust of the AMM during mission operations, tests should be conducted on mission-like simulations and field campaigns to test the operational limits of real-time decision making in mission relevant conditions and recognize any potential weaknesses that will need to be addressed prior to employing agentic AI facilitated by the AMM. The ultimate goal would be to make the AMM a robust flight-ready architecture capable of supporting diverse autonomous missions throughout the solar system both with and without humans in the loop.

### 5.3: Astrobiology-Chat (AB-Chat): A language interface for astrobiology integration

Astrobiology is a highly interdisciplinary field, requiring expertise in chemistry, biology, planetary science, and computational methods. However, the sheer volume of scientific literature and data from space missions poses a challenge for researchers to synthesize information efficiently. Astrobiological research demands extensive knowledge across diverse scientific literature and specialized understanding of multifaceted data from different traditional domains. For instance, the search for Martian biosignatures necessitates expertise spanning from prebiotic chemistry to geochemistry and microbial metabolisms, along with knowledge of the detailed interrogation of planetary data, such as imaging and spectral data from orbiters or rovers. This interdisciplinarity poses unique challenges; prebiotic chemistry papers are typically text-based, detailing experimental setups and chemical analyses, while planetary environment data, like spectral data, are often archived as numerical arrays requiring specialized software and knowledge of planetary science for interpretation. These differences in format and required expertise complicate the synthesis of information, identification of knowledge gaps, and communication across disciplinary boundaries.

**To address these challenges, it is suggested that a specialized text-centric large language model (LLM) incorporating interdisciplinary astrobiology knowledge is developed** (Figure 5)**.** This model would assist in identifying critical research gaps, facilitating hypothesis generation and testing, and supporting interactions with other models, representing a key instance of either a standalone FM or a downstream fine-tuning use case within a multimodal astrobiology FM.

Recognizing the value of domain-specific AI, NASA has already invested in the INDUS (a constellation name, stylized in caps) model; a suite of effective and efficient LLMs tailored for various scientific domains including planetary sciences and astrophysics (Bhattacharjee et al, 2024). The development of INDUS will also expose broad swaths of the science community to science based FMs, and improve adoption and acceptance of these tools and concepts. INDUS's encoders, based on the RoBERTa architecture, were trained on a substantial corpus of 60 billion tokens encompassing these relevant scientific fields. This domain-specific training provides INDUS with superior performance in understanding scientific language and concepts compared to general LLMs, making it a potentially suitable model for fine-tuning into a specialized astrobiology text-FM. Furthermore, an astrobiology-oriented LLM could serve as the interface to the multimodal use cases described above (i.e. life characterization and mission tasks) by assisting with prompt engineering for those models and supporting the utilization of outputs (see Section 1 and Figure 2). However, a standalone AB-Chat development effort (i.e. independent of a multimodal astrobiology FM) could proceed rapidly as a downstream application of a model such as the INDUS LLM, and several applications and initial steps are described in more detail here.

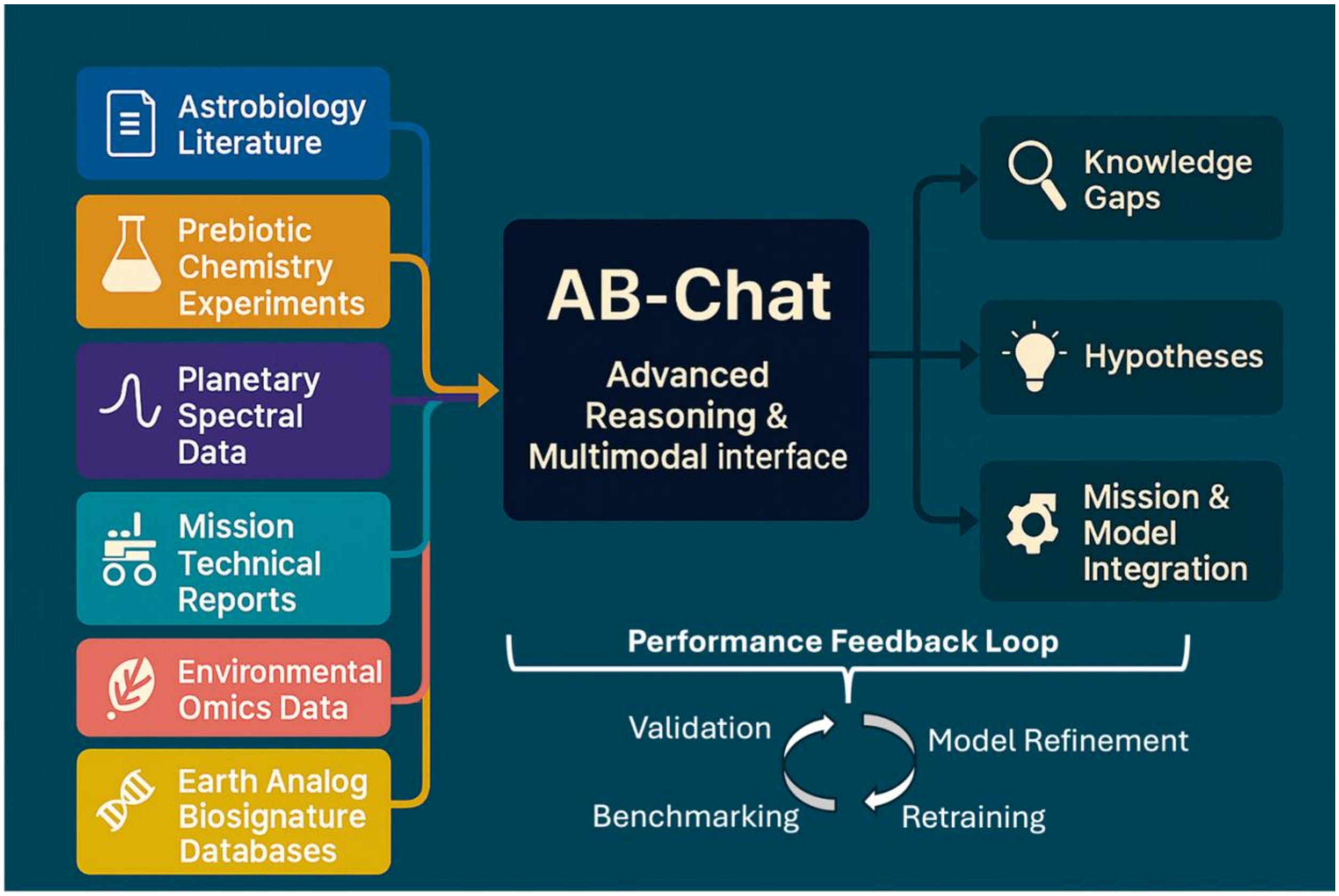

**Figure 5**: *An illustration of how a broad context AB-Chat would be trained and the roles it would serve to fill and augment.*

**AB-Chat applications**:
The potential applications of an AB-Chat model are extensive. First is the opportunity for systematic mining of astrobiology literature – including abstracts, peer-reviewed journals, and preprints. This activity could be extended to include deeper material from mission technical reports and documents, database user guides, and even metadata from dataset archives (in synergy with the AMM model, Section 5.2). AB-Chat could then be prompted to generate more terminology-agnostic search results and summaries. Generation of widely comprehensible content from technical and specialized literature can facilitate broader scientific communication, outreach, and student training. And mission and research development can be enhanced by synthesis of research or mission proposals and technical reports.

To truly elevate AB-Chat from a sophisticated search and summarization tool into an active, hypothesis-driven scientific collaborator, it will be essential to develop a dedicated, advanced reasoning module inspired by state-of-the-art systems such as OpenAI's o4 and Google's Gemini 2.5 Deep Research. This reasoning engine, developed entirely from the ground up, would integrate multi-step logical inference, chain-of-thought planning, and counterfactual evaluation capabilities vital for rigorously formulating and assessing astrobiological hypotheses. Coupled closely with retrieval-augmented generation (RAG) techniques from the fine-tuned INDUS-based language model, this reasoning component would enable AB-Chat to dynamically construct logical argument chains, evaluate alternative interpretations, and iteratively refine hypotheses as new planetary data and scientific findings become available. Thus, AB-Chat would not merely synthesize existing knowledge but actively drive the exploration and discovery processes fundamental to interdisciplinary astrobiology research.

**Potential AB-Chat development pathway stages**:

**(1): Data collection & curation:** Development would start with the further aggregation of astrobiologically relevant literature/data from key journals and additional resources not already captured by the INDUS training corpus (*e.g.,* adding focus on biosciences, origins of life), and mission and dataset-related documents. This effort would need to be combined with the development of structured ontologies and taxonomies (e.g., The Environment Ontology (ENVO), Chemical Entities of Biological Interest (ChEBI), Planetary Ontology, etc.) for critical astrobiology subdomains such as prebiotic chemistry, biosignatures, habitability, and planetary environments.

**(2): Model development & training:** Fine-tuning the LLM (that might be the multimodal astrobiology FM or an LLM such as INDUS) with curated astrobiology-specific datasets and implementing retrieval-augmented generation techniques to enhance knowledge recall and accuracy would be necessary. This effort would require the continuous validation and calibration of model outputs against expert-reviewed literature and validated scientific databases.

**(3): Full validation & benchmarking:** Model outputs would need to be tested against real-world research tasks - e.g., comparing hypotheses from Mars and Europa mission data, or interpreting biosignatures from planetary-analog environments on Earth. Performance would be assessed using benchmarks defined by expert review, domain-relevant metrics (e.g., cross-domain robustness), and stress-tested through adversarial queries to evaluate interpretability, citation accuracy, and epistemic reliability.

**(4): Deployment & expansion:** A functional AB-Chat model would be integrated into existing research infrastructures and databases (*e.g.*, NASA's Planetary Data System). User-centered design would enable hypothesis exploration, literature synthesis, and dynamic query interaction. Ongoing ingestion of new publications and datasets would support continual learning. Long-term maintenance would include version control, feedback incorporation, and expansion to support additional downstream tasks and data types as astrobiology research continues to evolve and grow.

AB-Chat would serve as a knowledge navigator, reducing the cognitive load by summarizing complex topics, identifying gaps in research, and facilitating hypothesis testing. As with all current LLMs and downstream models, it would not be expected to replace original human content. Rather, it will support and accelerate research by off-loading tasks. Just as modern search engines have largely replaced past practices for researching material by greatly increasing the efficiency of search and collation, an AB-Chat model would improve efficiency, understanding, and communication within the multidisciplinary astrobiology realm.

Existing general-purpose LLMs lack the specialized training necessary for astrobiology-specific tasks, making a domain-focused model essential for accelerating discoveries and improving collaboration across disciplines. By incorporating structured ontologies and validated datasets, AB-Chat would enhance the accuracy and relevance of AI-assisted research in astrobiology.

## 6. Summary, discussion, roadmap and vision

In this white paper, the participants of the Foundation Models for Astrobiology Workshop have presented findings for the Astrobiology community and its research programs. Three key areas (Section 5.1, 5.2, 5.3) have been identified where FMs could be used to advance the field and address astrobiology's central goal, to discover evidence of past or present life beyond Earth.

Throughout the in-person portion of the workshop the theme of data ecosystems and their roles in developing an FM or smaller ML tool repeatedly came up. This was first highlighted during the RCN presentations, while discussing how the RCNs handle disparate data across all of their member projects and what those data types are. The workshop participants came to the unanimous conclusion that, despite the rich *amount* of data within the astrobiology community, and NASA Open Science efforts, there is still a data identification and standardization barrier to combining large batches of multimodal data for unsupervised and supervised machine learning. Thus, data standardization is an essential next step for developing the FMs envisioned in this paper. Further details can be found in the report submitted to the DARES 2025 effort: *"The Astrobiology Data Ecosystem, Open Science, and the AI Era"* and a forthcoming white paper on Data Ecosystems. A high-level list of those identified needs are as follows.

1. Identifying, locating, and unifying all available astrobiology-related data
2. Obtaining needed data not currently available.
3. Improving access to unique resources (e.g. historical printed data).
4. Lowering barriers to implementation with streamlining and support.

Another emerging theme from the workshop was the overarching question of whether an astrobiology FM would be structured from a single multimodal FM architecture (see Figure 2) or might necessitate more than one architecture. For example, AB-Chat could potentially call for a specific LLM basis (such as INDUS that is based on BERT), while the data-orientated downstream use cases might call for a different architecture – although LLM FMs are capable of broad use with proper tokenization of data (e.g., GPT working with images or video). The workshop identified these questions and a need for a detailed technical evaluation of the capabilities of current FM architectures for the specific goals of astrobiology.

Furthermore, the three downstream use cases (Sections 5.1. 5.2, 5.3) that the workshop identified and focused on are likely only a subset of the potential use cases for an astrobiology FM or FMs. Other possibilities might include models specifically addressing functional biochemistry, environmental simulation, experimental design, astronaut safety, and planetary protection (forward/backward contamination). Additional work will need to be undertaken to develop a set of robust use cases and their priorities.

Figure 6 summarizes several key actions or goals that could form part of a roadmap and vision for an astrobiology FM effort that reflects the workshop findings. One ultimate, ambitious goal would be for an astrobiology FM (or FMs) to act as an effective "AI Astrobiologist" that would work alongside human investigators in any aspect of astrobiological research, potentially as an agentic AI.

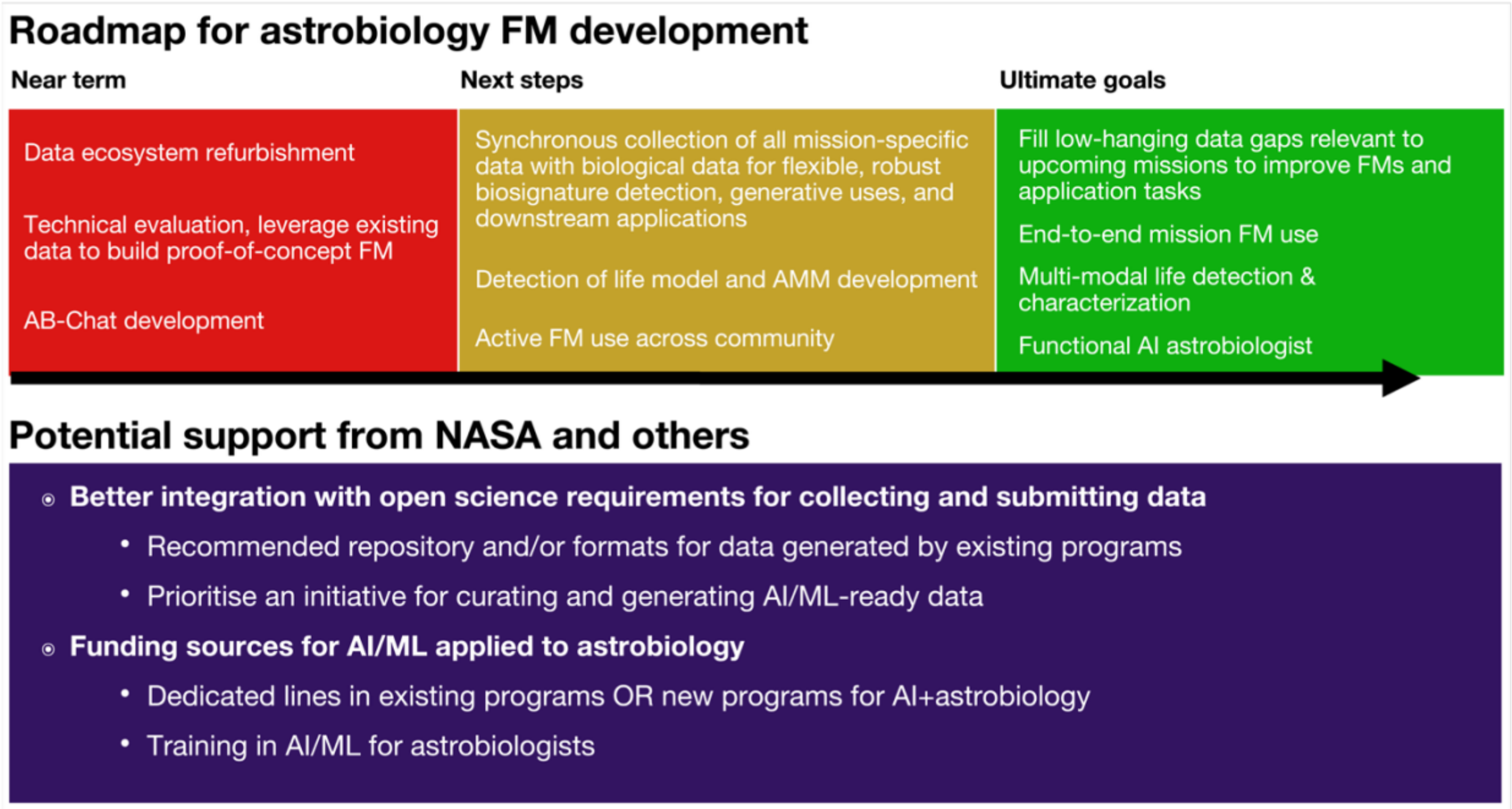

**Figure 6:** *Outline of workshop findings and recommendations for a roadmap towards astrobiology FM development and use, including key points for potential support areas from NASA and other sources.*

## Acknowledgements:

R. Felton and C. Scharf thank the NASA Astrobiology TWSC program (proposal# 24-TWSC24-0046) for its support of the FMs in Astrobiology Workshop and L. Hays, R. McCauley Rench, B. Burcar for their critical advice and direction, together with M. Voytek and M. Kirven-Brooks at NASA Ames for essential support and oversight. M. Ansdell is thanked for early discussions and advice. The SETI Institute provided the physical space to hold the workshop and supported the development of the program and a multitude of organizational tasks before, during, and after. Without the advice and on-hands hard work of Debbie Kolyer (SETI) the workshop would not have been a success, and we thank her and the President and CEO of the SETI Institute, Bill Diamond for their enormous generosity of time and attention. C. Nixon acknowledges funding support from NASA GSFC Strategic Science and NASA HQ Astrobiology. M. L. Wong is funded by NASA through the NASA Hubble Fellowship Program Grant HST-HF2-51521.001-A awarded by the Space Telescope Science Institute, which is operated by the Association of Universities for Research in Astronomy, Inc., for NASA, under contract NAS5-26555.

We also thank the essential contributions of the astrobiology RCNs, and specifically Tori Hoehler (NfOLD), Niki Parenteau (LIFE), Alison Murray (NOW), Loren Williams (PCE3), and Linda Sohl (NeXSS). Finally, we thank members of the Prithvi Geospatial FM team: Rahul Ramachandran (NASA-MSFC) and Manil Maskey (NASA-MSFC) for presenting to the workshop and their advice at many stages.

## Author Contribution Statement:

**Ryan Felton:** conceptualization (lead), funding acquisition (lead), project administration (lead), supervision (lead), visualization (lead), writing - original draft preparation (lead), writing - review and editing (lead). **Caleb Scharf:** conceptualization (lead), funding acquisition (lead), project administration (lead), supervision (lead), visualization (lead), writing - original draft preparation (lead), writing - review and editing (lead). **Stuart Bartlett:** conceptualization, writing - original draft preparation, writing - review and editing. **Nathalie A. Cabrol:** conceptualization (supporting), funding acquisition (supporting), project administration (supporting), supervision (supporting), visualization (supporting), writing - original draft preparation (supporting), writing - review and editing (supporting). **Victoria Da Poian:** conceptualization (supporting), project administration (supporting), visualization (supporting), writing - original draft preparation (supporting), writing - review and editing (supporting). **Diana Gentry:** conceptualization, visualization, writing - original draft preparation, writing - review and editing. **Jian Gong:** conceptualization, visualization, writing - original draft preparation, writing - review and editing. **Adrienne Hoarfrost:** conceptualization, visualization, writing - original draft preparation, writing - review and editing. **Manil Maskey:** conceptualization, visualization. **Floyd Nichols:**

conceptualization, visualization, writing - original draft preparation, writing - review and editing. **Conor A. Nixon:** conceptualization, visualization, writing - original draft preparation, writing - review and editing. **Tejas Panambur:** conceptualization, visualization, writing - original draft preparation, writing - review and editing. **Joseph Pasterski:** conceptualization, visualization, writing - original draft preparation, writing - review and editing. **Anton S. Petrov:** conceptualization, visualization, writing - original draft preparation, writing - review and editing. **Anirudh Prabhu:** conceptualization, visualization, writing - original draft preparation, writing - review and editing. **Brenda Thomson:** conceptualization, visualization, writing - original draft preparation, writing - review and editing. **Hamed Valizadegan:** conceptualization, visualization, writing - original draft preparation, writing - review and editing. **Kimberley Warren-Rhodes:** conceptualization, visualization, writing - original draft preparation, writing - review and editing. **David Wettergreen:** conceptualization (supporting), project administration (supporting), visualization (supporting), writing - original draft preparation (supporting), writing - review and editing (supporting). **Michael L. Wong:** conceptualization, visualization, writing - original draft preparation, writing - review and editing. **Anastasia Yanchilina:** conceptualization, visualization, writing - original draft preparation, writing - review and editing.

## Supplemental Information:

The workshop was managed by a six-member science organizing committee (SOC) made up of Ryan Felton, Caleb Scharf, Nathalie Cabrol, David Wettergreen, Victoria Da Poian, and Debbie Kolyer. The SOC distributed an application across multiple Astrobiology related listservs and to all five RCN's for them to disseminate amongst their members. Applicants responded to a questionnaire, submitted their CV's and were chosen based on career position, science background, and relevance of submitted material to the workshop. It was important to the SOC that a broad field of participants were chosen that spanned astrobiology, data science, and career level. The SOC ultimately chose 16 applicants (22 total including the SOC) that came from federal centers, universities, and private institutions and there were graduate students all the way up to career scientists in senior positions.

The workshop consisted of a 1-day virtual primer and 3-day in-person component. The full logistical schedule for the virtual primer and in-person are made available here in Supplemental File 1 and 2, and recordings from the workshop can be found at the following website: https://astrobio-fm2025.github.io/index.html.

## Glossary:

**Agentic AI** - AI systems that can plan and perform autonomous tasks with limited supervision, hence demonstrating a degree of independent decision-making and agency.

**Bespoke datasets** - Custom-built or highly domain-specific datasets.

**Curated data** - Data that has been selected, organized, and preprocessed by humans.

**Federated datasets** - Datasets sourced across multiple organizations but made accessible from one main hub.

**Few-shot learning** - A supervised machine learning technique where a model learns to make predictions from only a small number of labeled examples, sometimes just 2 or 3 from a category, when it is possible to infer from other contextual information.

**Foundation Model (FM)** - A semi-supervised trained model, typically trained on very large datasets, that serves as a general-purpose (hence “foundation”) base and can be adapted for a wide range of downstream tasks or applications.

**Large Language Models (LLMs)** - A type of foundation model trained on massive text datasets using deep learning to understand and generate natural language.

**Self-supervised learning** - A type of supervised machine learning for classification, where no labeled data is used for training, and the model generates their own labels or predictive tasks allowing them to extract meaningful patterns.

**Semi-supervised learning** - Machine learning approach that combines a small amount of labeled data with a large amount of unlabeled data during the training process to improve learning efficiency and cope with limited labeled data.

**Zero-shot learning** - A machine learning technique where the model is not provided with any labeled training data of a category and makes predictions on new tasks by transferring knowledge from pre-trained models of large, diverse datasets, for example recognizing a zebra on first encounter as a horse with stripes.

# Foundation Models in Astrobiology Workshop Agenda
# Feb 24 - 26, 2025

**Overarching goal**: To assess the potential of Foundation Models for astrobiology and to identify and investigate uses, model approaches, and what is needed for next steps, to support NASA and community decision-making and priorities.

*Coffee/drinks and light snacks provided throughout. Group lunches through Specialty's.*

## <u>Day 1 Monday Feb 24</u>: Theme - *Big Picture*

**9am**: Arrive at SETI and check-in
OC welcome

**9:20am**: Around the room introductions and icebreaker activities

**10am**: RCN presentations and discussions

**11:30am**: Bill Diamond (SETI) welcome speech

**12pm**: Lunch

**1pm**: Discussion of workshop goals, and identifying scientific focus areas of promise for FM development. Create list of fields and focus areas for FM's.

**1:30pm**: Breakout into groups aligned with areas from the list. Each group creates presentation to share with entire workshop:

- What would a given area's FM be used for – are there outstanding science questions suited to these models?
- What would the data for the FM be, and does the right data really exist?
- How could an FM really advance this area - is it necessary?
- What is the likely architecture and potential scale of this FM?
- If work on the FM began tomorrow what would that work be?

**2:15pm**: Break

**2:30pm**: Regroup for breakout teams to report findings and open discussion.

**4pm**: End Work – light reception until **5:30pm**

## <u>Day 2 Tuesday Feb 25</u>: Theme - *Deep Dive*

**9am**: Start - Examine priorities for work that have emerged.

**9:15am**: Decide on breakout groupings (same or revised) to dig deeper into Day 1 outcomes (i.e. specific FMs, use cases, priorities).

**9.30am**: Breakout groups perform deeper dives to flesh out the pathway to FMs for specific areas, determine data needs, and what a roadmap to these models would look like and anticipated use cases. Discuss multi-modality and scale of FM options.

- What would be required to build a specific FM?
- Can we outline the architecture and scale of potential FMs in more detail - if not, what is needed to do so?
- What is the level of challenge for utilizing or acquiring the necessary data for a given FM?
- What could be gained with multi-modal models? Is there a 'universal' FM for astrobiology?
- Where are the obvious hurdles or roadblocks to any of the above, and how might they be mitigated?

**10.45am**: Break

**11am**: Continue deep dive breakouts

**11:45am**: Group photo

**12pm**: Lunch

**1pm**: Continue deep dive breakouts

**2:30pm**: Break

**2:45pm**: Regroup for findings and discussions. Create slide deck summarizing findings at this point, to be used for debrief on Day 3 and as part of the planning process for whitepapers. OC will guide a recap of Day 3 tasks to steer towards those.

**5pm**: End Work

**7pm**: Optional group dinner at Don Giovanni, Mountain View

## <u>Day 3 Wednesday Feb 26</u>: Theme - *Writing & Debriefs*

**9am**: Check in and discussion of overall tasks for the day

**9:15am**: Share slide sets from Day 2. Group discussion of whitepaper(s)
Brainstorm on contents
Assign tasks, co-leads, and establish schedule
Begin writing

**10:45am**: Break

**11am**: Call with NASA HQ/Astrobiology leadership to debrief on workshop activities and Q&A

**12pm**: Lunch

**1pm**: Recap HQ discussions and discuss if there is a need to pivot on anything.

**1:15pm**: Resume writing. Create bulletized summaries of content for whitepaper(s), and begin to flesh out content.

**2.30pm**: Break and check-in

**2.45pm**: Continue as needed

**5pm**: Hard close.

**Virtual Primer Schedule - Wednesday Feb 5th, 2025**

**[15min talk / 15min discussion]**

**Start**: 8am PT / 11am ET

- Intro from OC - 8-8:15am
- Intro to unsup/sup/deep learning - *Floyd Nichols* - 8:15 - 8:45am
- ML pipeline - *Anton Petrov* - 8:45 - 9:15am
- Ethics/interpretability - *Victoria Da Poian* - 9:15 - 9:45am

**Break:** 9:45 - 10:00am

- Foundation Models - *Adrienne Hoarfrost* - 10:00 - 10:30
- Biosignatures - *Stuart Bartlett* - 10:30 - 11:00
- Exoplanet characterization/habitability - *Mike Wong* - 11:00 - 11:30

**Break:** 11:30 - 11:45

- Sample composition - *Joseph Pasterski* - 11:45 - 12:15
- In-situ autonomy - *Open group discussion* - 12:15 - 12:45

**End:** 12:45pm PT /3:45pm ET